\documentclass{JHEP3}
\usepackage{cite,amsmath,times,relsize}
\newcommand{\north}{\text{\smaller[3]N}}
\newcommand{\south}{\text{\smaller[3]S}}
\newcommand{\rch}{q}
\newcommand{\nin}{ \;/\!\!\!\!\in }
\newcommand{\susy}{{\bf Q}}
\newcommand{\brs}{{\bf Q}_\text{B}}
\newcommand{\sdot}{{\cdot}}
\newcommand{\RV}{{\text{R}_\text{V}}}
\newcommand{\RA}{{\text{R}_\text{A}}}

\title{Orbifolds, Defects and Sphere Partition Function}

\author{
Kazuo Hosomichi$^{a,b}$\\
$^a$Department of Physics, National Taiwan University, Taipei 10617, Taiwan\\
$^b$Physics Division, National Center for Theoretical Sciences\\
~National Tsing-Hua University, Hsinchu 30013, Taiwan
\vskip2mm
E-mail:
\email{hosomiti@phys.ntu.edu.tw}
}

\abstract{Gauge theories in the presence of codimension two vortex
defects are known to be related to the theories on orbifolds. By using
this relation we study the localized path integrals of 2D ${\cal N}=(2,2)$
SUSY gauge theories with point-like vortex defects. We present a formula
for the correlation functions of vortex defects inserted at the north and the
south poles of squashed spheres. For Abelian gauge theories the
correlators are locally constant as functions of the parameters of the
defect, but exhibit discontinuity at some threshold values determined
by the deficit angle at the poles and the R-charges of the matter multiplets. For non-Abelian gauge
groups the correlators depend non-trivially on the types of gauge
symmetry breaking due to the defects.
}

\preprint{NCTS-TH/1503}

\begin{document}

\section{Introduction}\label{sec:intro}

In quantum field theories in different dimensions, there are interesting
class of operators which are not defined in the ordinary way as
functionals of fields. These operators, sometimes called defects, are
defined by requiring specific singular behavior on the fields around them.
One example is the 't Hooft operators \cite{'tHooft:1977hy} in 4D gauge
theories supported along one-dimensional paths, whose definition
involves requiring the gauge field to develop Dirac monopole singularity
along the paths. Another example is the vortex defects supported on
codimension two submanifolds. For 4D gauge theories, they are supported
along 2D surfaces and therefore called surface defects. They are defined
by the singular behavior of the gauge field $A$
\begin{equation}
 A\sim \eta\cdot d\varphi
\label{defvd}
\end{equation}
near the defect. Here $\eta$ is an element of the Lie algebra, and
$\varphi$ is the angular coordinate on the 2D plane transverse to and
centered at the defect. See \cite{Gukov:2014gja} for a review on
the recent developments. Similar definition in 3D supersymmetric gauge
theories leads to 1D defects which are sometimes called vortex loops.

These defects will allow us to study new aspects of quantum field
theories related to monopoles or other solitons carrying topological
charges in gauge theories. It is particularly interesting to study such
defects in the classes of supersymmetric theories in 2, 3 or 4 dimensions
where various dualities or nontrivial relations among observables are known.
For example, the inclusion of surface defects in 4D ${\cal N}=2$ supersymmetric
gauge theories has been studied in the computation of instanton
partition functions
\cite{Kozcaz:2010af,Dimofte:2010tz,Taki:2010bj,Awata:2010bz,Kozcaz:2010yp,Wyllard:2010rp,Wyllard:2010vi,Bonelli:2011fq,Kanno:2011fw,Bonelli:2011wx},
superconformal indices
\cite{Nakayama:2011pa,Gaiotto:2012xa,Alday:2013kda,Gadde:2013ftv,Bullimore:2014awa} 
or sphere partition functions
\cite{Gomis:2014eya,Nawata:2014nca,Gaiotto:2014ina,Chen:2015fta},
and the results led to a more detailed understanding of the relation
between 4D ${\cal N}=2$ SUSY theories and 2D conformal field theories
\cite{Alday:2009fs,Drukker:2009id,Frenkel:2015rda}, topological field
theories or topological strings. The loop operators in 4D ${\cal N}=2$
theories were studied from a similar viewpoint in
\cite{Gomis:2010kv,Gomis:2011pf,Gang:2012yr}; see \cite{Okuda:2014fja}
for a review.
Another interesting point is that some of the defect operators can also
be described alternatively by a lower-dimensional field theory on their
worldvolume interacting with the fields in the bulk
\cite{Gomis:2014eya,Assel:2015oxa}.

A powerful computational tool to evaluate supersymmetric observables
explicitly is the localization principle. To use this for supersymmetric
path integrals with defect observables, we need to know how to perform
the path integration with modified boundary condition on fields. For 't
Hooft loop operators in 4D ${\cal N}=2$ supersymmetric gauge theories
this was studied in \cite{Gomis:2011pf,Ito:2011ea}, where a result
fully consistent with the prediction of S-duality was recovered by a
careful localization analysis. Similar analysis was made for vortex loop
operators (\ref{defvd}) in 3D SUSY gauge theories in
\cite{Kapustin:2012iw,Drukker:2012sr}.

The purpose of this paper is to study the codimension two vortex defects
of the type (\ref{defvd}) in 2D ${\cal N}=(2,2)$ SUSY gauge theories,
which are actually local
operators. We study the defects by putting two of them at the north and
the south poles of the sphere $S^2$ and evaluating their correlation
function explicitly. The key idea in our analysis is the equivalence of
the gauge theories in the presence of the defect and the gauge theories on
orbifolds \cite{Biswas:1997aa}. This equivalence was relevant in the
description of the moduli space of instantons in the surface defect
background and the computation of the so-called ramified instanton partition
function, see \cite{Kanno:2011fw} and the references therein. Based on
the same idea, a formula for the $S^4$ partition function of
${\cal N}=2$ SUSY gauge theories in the presence of a surface defect was
proposed in \cite{Nawata:2014nca}.

We study the defect correlators by reducing the path integral to a
finite-dimensional integral and sum using the idea of Coulomb branch
localization. The integrand then depends on the parameter $\eta$
of the vortex defects (\ref{defvd}). When the gauge group is Abelian, it
turns out that the dependence on $\eta$ can be completely absorbed into
the redefinition of the remaining integration variables when
$\eta$ is within a certain range. As a function of $\eta$ the defect
correlator is therefore locally constant, but there are threshold values
of $\eta$ at which the value of correlators jumps. The detail of this
behavior is shown in the example of $U(1)$ gauge theory (SQED) with
$N_\text{F}$ electrons and $N_\text{A}$ positrons. For non-Abelian gauge
theories, the above discontinuous behavior of the integrand leads to a
non-trivial dependence of the defect correlator on the type of gauge
symmetry breaking at the poles, characterized by the Levi subgroups. We
illustrate this by calculating a few sample defect correlators in the
$U(N)$ SQCD with $N_\text{F}$ fundamental and $N_\text{A}$
anti-fundamental chiral multiplets.

The organization of this paper is as follows. Section \ref{sec:theory}
is a review of the construction of 2D ${\cal N}=(2,2)$ SUSY gauge
theories on squashed spheres, in which we set up the necessary
conventions. We will restrict ourselves to the theories of vector and
chiral multiplets. In Section \ref{sec:PF} we revisit the computation of
exact partition functions using the index theorem and the fixed point
formula. In Section \ref{sec:defect} we study the localized path
integral on the defect background by making connection with the
partition functions on orbifolds of $S^2$, and propose a formula for the
defect correlators. We conclude with a few remarks in Section \ref{sec:concl}.

\paragraph{Notations.}

We use Pauli's matrices as the 2D gamma matrices $\gamma^1,\gamma^2$ and
the chirality matrix $\gamma^3$. Their antisymmetrized products are
denoted as $\gamma^{ab}=\gamma^{[a}\gamma^{b]}$. The charge conjugation
matrix $\epsilon$ is anti-symmetric and satisfies
$\epsilon\gamma^a\epsilon^{-1}=-(\gamma^a)^T$. Explicitly,
\begin{equation}
\def\arraystretch{.85}
 \epsilon=\left(\begin{array}{cc} 0&1  \\ -1& 0 \end{array}\right),\quad
 \gamma^1=\left(\begin{array}{cc} 0&1  \\  1& 0 \end{array}\right),\quad
 \gamma^2=\left(\begin{array}{cc} 0&-i \\  i& 0 \end{array}\right),\quad
 \gamma^3=\left(\begin{array}{cc} 1&0  \\  0& -1\end{array}\right).
\def\arraystretch{1.0}
\end{equation}
For bilinear products of spinors we use the short-hand notations,
\begin{equation}
 \xi\psi \equiv \xi^\alpha\epsilon_{\alpha\beta}\psi^\beta,\quad
 \xi\gamma^a\psi \equiv \xi^\alpha
 \epsilon_{\alpha\beta}(\gamma^a)^\beta_{~\gamma}\psi^\gamma,\quad
\text{etc.}
\end{equation}
Note that $\epsilon_{\alpha\beta}$ is antisymmetric while
$(\epsilon\gamma^a)_{\alpha\beta}$ are all symmetric.

\section{Rigid SUSY on curved 2D surfaces}\label{sec:theory}

We review here the construction of a class of ${\cal N}=(2,2)$ SUSY
gauge theories on curved backgrounds. The results summarized here were
obtained by \cite{Benini:2012ui,Doroud:2012xw} for the SUSY gauge
theories on round sphere $S^2$, and later generalized by \cite{Gomis:2012wy}
to squashed spheres.

\paragraph{Killing spinors.}

We consider two-dimensional spaces with Killing spinors $\xi,\bar\xi$
satisfying
\begin{eqnarray}
 D_m\xi &\equiv&
 \left(\partial_m+\frac14\omega_m^{ab}\gamma^{ab}-iV_m\right)\xi
 ~=~ \gamma_m\xi'\,,
 \nonumber \\
 D_m\bar\xi &\equiv&
 \left(\partial_m+\frac14\omega_m^{ab}\gamma^{ab}+iV_m\right)\bar\xi
 ~=~ \gamma_m\bar\xi'\,~~
 \text{for some }\xi',\bar\xi'.
\label{KSeq}
\end{eqnarray}
Here $V_m$ is the background vector field which gauges the vector $U(1)$
R-symmetry $\RV$. So the spinor $\xi$ carries the R-charge
$\RV=+1$ and $\bar\xi$ carries $\RV=-1$. Throughout this paper we take
the spinors $\xi,\bar\xi$ to be Grassmann even. Most of the time we
focus on backgrounds on which the Killing spinors satisfy
\begin{equation}
 D_m\xi = \frac{iH}{2}\gamma_m\xi,\quad
 D_m\bar\xi = \frac{iH}{2}\gamma_m\bar\xi\,,
\label{KSeq2}
\end{equation}
where $H$ a scalar auxiliary field.

An example of supersymmetric curved backgrounds is the round sphere
$S^2$ of radius $\ell$ with the background fields $V_m=0, H=1/\ell$. The
vielbein and spin connection are written in polar coordinates as
\begin{equation}
 e^1 = \ell d\theta,\quad
 e^2 = \ell \sin\theta d\varphi,\quad
 \omega^{12} = -\cos\theta d\varphi\,.
\end{equation}
The Killing spinor equation (\ref{KSeq2}) then has a solution
\begin{equation}
 \xi = e^{-\frac{i\varphi}2}
 \left(\begin{array}{r}
 i\sin\frac\theta2 \\
  \cos\frac\theta2 \end{array}\right),\quad
 \bar\xi = e^{\frac{i\varphi}2}
 \left(\begin{array}{r}
  \cos\frac\theta2 \\
 i\sin\frac\theta2 \end{array}\right).
\label{KS}
\end{equation}
As has been shown in \cite{Gomis:2012wy}, one can squash the sphere
into an ellipsoid of axis-lengths $\ell,\ell,\tilde\ell$ embedded in
$\mathbb R^3$ preserving the supersymmetry corresponding to the above
$\xi,\bar\xi$. The natural choice of vielbein
and the corresponding spin connection are
\begin{equation}
 e^1=f(\theta) d\theta, \quad
 e^2=\ell\sin\theta d\varphi,\quad
 \omega^{12}=-\frac{\ell}{f(\theta)}\cos\theta d\varphi,
\label{BG1}
\end{equation}
with $f(\theta) \equiv \sqrt{\ell^2\cos^2\theta+\tilde\ell^2\sin^2\theta}$.
The spinor fields (\ref{KS}) satisfy (\ref{KSeq}) if the background
fields are chosen as
\begin{equation}
 H = \frac1f,\quad
 V = \frac12\left(\frac{\ell}{f}-1\right)d\varphi\,.
\label{BG2}
\end{equation}
In fact, the background defined by (\ref{BG1}), (\ref{BG2}) admits
the Killing spinors (\ref{KS}) for arbitrary choice of the function $f(\theta)$.

\paragraph{Construction of SUSY gauge theories.}

In this paper we restrict our attention to the ${\cal N}=(2,2)$
supersymmetric gauge theories of vector and chiral multiplets.
The vector multiplet consists of the Lie algebra valued fields -- the
gauge field $A_m$, two real scalars $\rho,\sigma$, spinors
$\lambda,\bar\lambda$ and the auxiliary scalar field $D$. They transform
under supersymmetry as
\begin{eqnarray}
 \susy A_m &=& \tfrac12(\xi\gamma_m\bar\lambda+\bar\xi\gamma_m\lambda),
 \nonumber \\
 \susy\rho &=& \tfrac12(\xi\gamma_3\bar\lambda+\bar\xi\gamma_3\lambda),
 \nonumber \\
 \susy\sigma &=& \tfrac i2(\xi\bar\lambda-\bar\xi\lambda),
 \nonumber \\
 \susy\lambda &=& \tfrac12\gamma^{mn}\xi F_{mn}
  -\gamma^{3m}D_m(\xi\rho)-i\gamma^mD_m(\xi\sigma)
  -\gamma^3\xi[\rho,\sigma]+\xi D,
 \nonumber \\
 \susy\bar\lambda &=& \tfrac12\gamma^{mn}\bar\xi F_{mn}
 -\gamma^{3m}D_m(\bar\xi\rho)+i\gamma^mD_m(\bar\xi\sigma)
 +\gamma^3\xi[\rho,\sigma]-\bar\xi D,
 \nonumber \\
 \susy D &=& \tfrac12D_m(\xi\gamma^m\bar\lambda-\bar\xi\gamma^m\lambda)
 -\tfrac i2[\rho,\xi\gamma_3\bar\lambda-\bar\xi\gamma_3\lambda]
 +\tfrac 12[\sigma,\xi\bar\lambda-\bar\xi\lambda]\,.
\end{eqnarray}
A chiral multiplet consists of a scalar $\phi$, a spinor $\psi$
and an auxiliary scalar $F$ in a complex representation $\Lambda$ of the
gauge group. The multiplet is labeled by the vector R-charge of the lowest
component,
\begin{equation}
 \RV[\phi]=2\rch.
\end{equation}
The conjugate fields $\bar\phi,\bar\psi,\bar F$ form an anti-chiral
multiplet in the representation $\bar{\Lambda}$ of the gauge group. The
supersymmetry acts on the fields as
\begin{eqnarray}
 \susy\phi &=& \xi\psi,
 \nonumber \\
 \susy\bar\phi &=& \bar\xi\bar\psi,
 \nonumber \\
 \susy\psi &=& -\gamma^m\bar\xi D_m\phi
 +i\gamma^3\bar\xi\rho\phi-\bar\xi\sigma\phi
 -\rch\cdot\phi\gamma^mD_m\bar\xi+\xi F,
 \nonumber \\
 \susy\bar\psi &=& -\gamma^m\xi D_m\bar\phi
 -i\gamma^3\xi\bar\phi\rho-\xi\bar\phi\sigma
 -\rch\cdot\bar\phi\gamma^mD_m\xi-\bar\xi\bar F,
 \nonumber \\
 \susy F &=& -\bar\xi(\gamma^mD_m\psi
 -i\gamma^3\rho\psi-\sigma\psi-i\bar\lambda\phi)
 -\rch\cdot D_m\bar\xi\gamma^m\psi,
 \nonumber \\
 \susy\bar F &=& +\bar\xi(\gamma^mD_m\bar\psi
 +i\gamma^3\bar\psi\rho-\bar\psi\sigma+i\bar\phi\lambda)
 +\rch\cdot D_m\xi\gamma^m\bar\psi.
\end{eqnarray}
Here the fields belonging to the representation $\Lambda$
($\bar\Lambda$) of the gauge group is regarded as column vectors 
(resp. row vectors), on which the vector multiplet fields act from the
left (resp. right).

The supersymmetry $\susy$ squares into a sum of bosonic symmetries,
\begin{eqnarray}
&&
 \susy^2 ~=~ \text{Lie}(v)+\text{Lorentz}(\theta_{ab})+\text{Weyl}(\omega)
 +\text{Gauge}(\Sigma)+\RV(\alpha)+\RA(\beta),
 \nonumber \\ 
&&
 \text{\footnotesize (example)}~~
 \susy^2\lambda ~=~
 v^m\partial_m\lambda+\frac14\theta_{ab}\gamma^{ab}\lambda
 +\frac32\omega\lambda
 +[\Sigma,\lambda]+\alpha\lambda+\beta\gamma_3\lambda,
\end{eqnarray}
with the transformation parameters $v^m=-\bar\xi\gamma^m\xi$ and
\begin{eqnarray}
 \theta_{ab} &=& e_a^me_b^nD_{[m}v_{n]} + v^m\omega_{m,ab},
 \nonumber \\
 \omega &=& \tfrac12 D_mv^m,
 \nonumber \\
 \Sigma &=& \bar\xi\xi\sigma+i\bar\xi\gamma_3\xi\rho-iv^m A_m,
 \nonumber \\
 \alpha &=& \tfrac14(\bar\xi\gamma^mD_m\xi-D_m\bar\xi\gamma^m\xi)-iv^mV_m,
 \nonumber \\
 \beta  &=& \tfrac14(\bar\xi\gamma^{m3}D_m\xi-D_m\bar\xi\gamma^{m3}\xi).
\end{eqnarray}
Note that the transformation rules are invariant under Weyl rescalings
which transform the metric as a weight $-2$ object,
$\tilde g_{mn}=g_{mn}e^{2\omega}$.
The scaling weights, $\RV$ and the $\RA$-charges of the fields are
summarized in the Table \ref{tab:weyl-R}.
\begin{table}
\def\STRUT{\rule[-0.85mm]{0mm}{5.5mm}}
{\small
\begin{center}
\begin{tabular}{c||cc|ccccc|cccccc}
\hline
\!\!fields & $\xi$ & $\bar\xi$ &
 $A_m$ & $\rho\pm i\sigma$ & $\lambda$ & $\bar\lambda$ & $D$ &
 $\phi$ & $\bar\phi$ & $\psi$ & $\bar\psi$ & $F$ & $\bar F$ \STRUT\\
\hline
\!\!scale & $-\frac12$ & $-\frac12$ &
 $0$ & $1$ & $\frac32$ & $\frac32$ & $2$ &
 $\rch$ \!\!\!&\!\!\! $\rch$ \!\!\!&\!\!\! $\rch+\frac12$ \!\!\!&\!\!\!
 $\rch+\frac12$ \!\!\!&\!\!\! $\rch+1$ \!\!\!&\!\!\! $\rch+1$\!\! \STRUT\\
\hline
$\text{R}_{\text{V}}$ & $1$ & $-1$ &
 $0$ & $0$ & $1$ & $-1$ & $0$ &
 $2\rch$ \!\!\!&\!\!\! $-2\rch$ \!\!\!&\!\!\! $2\rch-1$ \!\!\!&\!\!\!
 $1-2\rch$ \!\!\!&\!\!\! $2\rch-2$ \!\!\!&\!\!\! $2-2\rch$\!\! \STRUT\\
\hline
$\text{R}_{\text{A}}$ & $\gamma_3$ & $-\gamma_3$ &
 $0$ & $\pm2$ & $\gamma_3$ & $-\gamma_3$ & $0$ &
 $0$ & $0$ & $\gamma_3$ & $-\gamma_3$ & $0$ & $0$ \STRUT\\
\hline
\end{tabular}
\label{tab:weyl-R}
\caption{the scaling weight and the R-charge of the fields.}
\end{center}
}
\end{table}

As the kinetic terms for these fields one can take
\begin{eqnarray}
 {\cal L}_{\text{vec}} &=&
 \text{Tr}\Big[
 \big(F_{12}-H\rho\big)^2
 +\big(D+H\sigma\big)^2+D_m\rho D^m\rho+D_m\sigma D^m\sigma
 -[\rho,\sigma]^2
  \nonumber \\ && \hskip6mm
 -\bar\lambda\big(
  \gamma^mD_m\lambda-i\gamma^3[\rho,\lambda]-[\sigma,\lambda]\big)
  \Big],
 \nonumber \\
 {\cal L}_{\text{mat}} &=&
 D_m\bar\phi D^m\phi+\bar\phi\Big\{\rho^2+\sigma^2+2i\rch H\sigma
 +\tfrac12\rch R-\rch^2H^2+iD\Big\}\phi
 +\bar FF
 \nonumber \\ &&
 -\bar\psi\Big(\gamma^mD_m-i\gamma^3\rho-\sigma-i\rch H\Big)\psi
 +i\bar\psi\bar\lambda\phi-i\bar\phi\lambda\psi\,.
\end{eqnarray}
These are invariant under the SUSY generated by the Killing spinors
satisfying the stronger condition (\ref{KSeq2}). They are in fact SUSY
exact,
\begin{eqnarray}
 {\cal L}_\text{vec} &=& \susy_\xi\susy_{\bar\xi}\text{Tr}\left[
  \bar\lambda\lambda+4i\sigma D+2iH\sigma^2 \right],
 \nonumber \\
 {\cal L}_\text{mat} &=& \susy_\xi\susy_{\bar\xi}\left[
  -\bar\psi\psi-2\bar\phi\sigma\phi-i(2\rch-1)H\bar\phi\phi \right],
\end{eqnarray}
so they play the role of regulators.
Namely, since the values of SUSY path integrals do not depend on the
coefficients multiplying them, one can make them large so that the saddle
point approximation becomes exact and the path integral localizes onto
the moduli space of $\susy$-invariant configurations. On the other hand,
the FI-theta term for Abelian vector multiplet,
\begin{equation}
 {\cal L}_{\text{FI}}~=~-i\zeta D + \frac{i\theta}{2\pi}F_{12},
\end{equation}
is SUSY invariant but not exact, so the expectation values of
supersymmetric observables depend non-trivially on the couplings
$\zeta,\theta$. Another coupling is the mass for chiral multiplets,
which can be introduced by coupling some external vector multiplets to
the matter flavor symmetries and turning on its scalar components, as
we will see later.

\section{Ellipsoid partition function}\label{sec:PF}

The exact formula for partition functions on the ellipsoid (\ref{BG1})
has been derived by \cite{Gomis:2012wy}, generalizing the earlier result
for the round sphere \cite{Benini:2012ui,Doroud:2012xw}. We revisit here
the derivation of the formula using the index theorem.

Using Coulomb branch localization, the path integral can be shown to 
localize onto the saddle point configurations which minimize the bosonic
part of ${\cal L}_\text{vec}$. The saddle points are labeled by the
parameters $s,a$ taking values in the Cartan subalgebra. Up to gauge
choices, the bosonic fields in vector multiplet take the following form
\begin{equation}
 \sigma = \frac a\ell,\quad
 D = -\frac a{f\ell},\quad
 A = s\cdot\cos\theta d\varphi,\quad
 \rho = -\frac s\ell.
\label{saddle}
\end{equation}
while the matter fields all vanish.
The parameter $s$ is the magnetic flux through the ellipsoid, so it is
subject to GNO quantization. The path integral thus decomposes
into two steps: one first integrates over the fluctuations around
each saddle point under Gaussian approximation and obtain the so-called
one-loop determinant. The result is then summed (integrated) over all
the saddle points.

\paragraph{From determinant to character.}

The one-loop determinant can be most easily computed by moving to a new set of
path integration variables. Let us explain this change of variables
for the chiral multiplet first. For simplicity we consider a chiral multiplet
with unit charge under a $U(1)$ vector multiplet which take the
saddle point value (\ref{saddle}) with $a\in\mathbb{R}$, $s\in\mathbb{Z}/2$.
From the fields $\phi,\psi, F$ we define a boson $\bf X$, a fermion
$\boldsymbol\Xi$ and their superpartners $\susy{\bf X}$,
$\susy\boldsymbol\Xi$ as follows,
\begin{equation}
\begin{array}{l}
 {\bf X}\equiv\phi, \\
 \bar{\bf X}\equiv\bar\phi,
\end{array}
~~
\begin{array}{l}
 \susy{\bf X}\equiv\xi\psi,\\
 \susy\bar{\bf X}\equiv\bar\xi\bar\psi,
\end{array}
~~
\begin{array}{l}
 {\boldsymbol\Xi}\equiv\bar\xi\psi,\\
 \bar{\boldsymbol\Xi}\equiv\xi\bar\psi,
\end{array}
~~
\begin{array}{l}
 \susy{\boldsymbol\Xi}
 =F-\bar\xi\gamma^m\bar\xi D_m\phi+i\bar\xi\gamma^3\bar\xi\rho\phi,\\
 \susy\bar{\boldsymbol\Xi}
 =\bar F-\xi\gamma^m\xi D_m\bar\phi-i\xi\gamma^3\xi\bar\phi\rho\,.
\end{array}
\end{equation}
Note that ${\bf X}$ is Grassmann-even while ${\boldsymbol\Xi}$ is odd,
but they are both Lorentz scalars.
One can show that the relation between $\phi,\psi,F$ and the new fields
${\bf X},\susy{\bf X}, {\boldsymbol\Xi},\susy{\boldsymbol\Xi}$ is local
and invertible.
The path integral over the fluctuations then gives rise to the ratio
of determinants of $\susy^2$ acting on the fields ${\bf X}$ and
$\boldsymbol\Xi$,
\begin{equation}
 Z_\text{1-loop}=\frac{\text{det}\susy^2|_{\boldsymbol\Xi}}
                  {\text{det}\susy^2|_{\bf X}}\,.
\label{Z1l}
\end{equation}
This can be understood by thinking of the $\susy$-exact path
integration weight $e^{-S_\text{reg}}=e^{-\susy{\cal V}}$ with
\begin{eqnarray}
 {\cal V} &=&
 \int d^2x\sqrt{g}\big(
 \bar{\bf X}\cdot\susy{\bf X}+
 \bar{\boldsymbol\Xi}\cdot\susy{\boldsymbol\Xi}\big)\,,
 \nonumber \\
S_\text{reg} &=&
 \int d^2x\sqrt{g}\big(
 \susy\bar{\bf X}\cdot\susy{\bf X}
+\bar{\bf X}\cdot\susy^2{\bf X}
+\susy\bar{\boldsymbol\Xi}\cdot\susy{\boldsymbol\Xi}
-\bar{\boldsymbol\Xi}\cdot\susy^2{\boldsymbol\Xi}\big)\,.
\end{eqnarray}
The above choice of $S_\text{reg}$ is the simplest, but does not
necessarily have a positive definite bosonic part, so the Wick rotation
of some integration variables will be needed. More careful analysis with
an $S_\text{reg}$ with manifestly positive definite bosonic part should
lead to the same one-loop determinant.

As a useful intermediate quantity for computing the one-loop determinant
(\ref{Z1l}), let us introduce the character
\begin{equation}
 \chi = \text{Tr}_{\bf X}(e^{t\susy^2})
-\text{Tr}_{\boldsymbol\Xi}(e^{t\susy^2}).
\label{ch}
\end{equation}
The fields ${\bf X}$ and ${\boldsymbol\Xi}$ are scalars with the vector
R-charge $2\rch$ and $2\rch-2$. The square of the SUSY acts on
scalar fields in general as
\begin{equation}
 \susy^2 = \frac1\ell\left\{
 iJ_3 +\frac i2\RV +\text{Gauge}(a)
 \right\},\quad J_3\equiv -i\partial_\varphi \,.
\end{equation}
The character (\ref{ch}) can be computed as an index if there is a
differential operator which commutes with $\susy^2$ and maps between
${\bf X}$ and ${\boldsymbol\Xi}$. Such operators $J^\pm$ with the vector
R-charge $\mp2$ can be constructed using Killing spinors,
\begin{equation}
 J^+ = \ell\left(
 \bar\xi\gamma^m\bar\xi D_m-i\bar\xi\gamma^3\bar\xi\rho \right),
 \quad
 J^- = \ell\left(
 \xi\gamma^m\xi D_m-i\xi\gamma^3\xi\rho \right).
\end{equation}
More explicitly in components,
\begin{equation}
 J^\pm = e^{\pm i\varphi}\left\{
 \pm\frac\ell f\partial_\theta+i\cot\theta\partial_\varphi
 +\frac s{\sin\theta}+\frac12\RV\Big(\frac\ell f-1\Big)\cot\theta
 \right\}.
\label{Jpm}
\end{equation}
Note that when $f(\theta)=\ell$ these operators coincide with the
angular momentum operators in the background magnetic flux $s$,
justifying the name $J^\pm$. Therefore, on the round sphere the
character can be computed by expanding the fields ${\bf X}$ and
${\boldsymbol\Xi}$ into monopole harmonics \cite{Wu:1976ge}.

Monopole harmonics is a convenient basis to expand the fields on $S^2$
coupled to the background monopole gauge field
\begin{equation}
 A= s\cos\theta d\varphi,\quad
 F=-s\sin\theta d\theta d\varphi = -\frac s{2r^3}\epsilon_{abc}x_adx_bdx_c,
\quad
 (r\equiv\sqrt{x_ax_a})
\label{u1gauge}
\end{equation}
where $x_a$ is the Cartesian coordinate of the embedding space $\mathbb R^3$.
The angular momentum operators for a particle with unit electric charge is
\begin{eqnarray}
&&
J_a^{(s)}=-i\epsilon_{abc}x_b(\partial_c-iA_c)+\frac{sx_a}r,
\nonumber \\ &&
 J_3^{(s)}=-i\partial_\varphi,\quad
 J_\pm^{(s)}=e^{\pm i\varphi}
 \left(\pm\partial_\theta+i\cot\theta\partial_\varphi
 +\frac s{\sin\theta}\right).
\end{eqnarray}
Note that $J_\pm^{(s)}$ agree with the differential operators
(\ref{Jpm}) with $f(\theta)=\ell$ substituted. 
The monopole harmonics $Y^s_{jm}$ transform as the
spin-$j$ representation under the action of $J_a^{(s)}$. Let us also
introduce
\begin{equation}
 D_\pm^{(s)}\equiv \partial_\theta \mp \frac i{\sin\theta}
 (\partial_\varphi-is\cos\theta).
\label{Dirac}
\end{equation}
One can show that the massless Dirac operator for the charged Weyl
spinors of $\pm$ chirality on sphere is $D_\mp^{(s\pm1/2)}$, where the
$\pm1/2$ shift is due to spin connection. Note also that the Hermite
conjugate of $D_\pm^{(s)}$ is $D_\mp^{(s\pm1)}$, since the volume
element of the unit sphere is $\sin\theta d\theta d\varphi$.
It now follows from the important identities
\begin{eqnarray}
 D_\pm^{(s)}J_a^{(s)} &=&  J_a^{(s\pm1)}D_\pm^{(s)},
 \nonumber \\
 D_\pm^{(s\mp1)}D_\mp^{(s)} &=& s(s\mp1)-J_a^{(s)}J_a^{(s)},
\end{eqnarray}
that $Y_{jm}^s$ exists only for $j\in |s|+\mathbb Z_{\ge0}$. One can
normalize the monopole harmonics so that the following holds,
\begin{equation}
 D_\pm^{(s)}Y^s_{jm}= \pm\sqrt{(j\mp s)(j\pm s+1)}Y^{(s\pm1)}_{jm}.
\label{D-sph}
\end{equation}

The character on the round sphere can be easily evaluated by recalling
that the kernel of $J^+$ (or $J^-$) are spanned by the monopole
harmonics $Y^s_{jm}$ with $m=j$ (resp. $m=-j$), and that the allowed
values of $j$ are bounded by $|s|$.
\begin{eqnarray}
 \chi(a,s) &=&
 \text{Tr}_{\bf X}\big(e^{t\susy^2}\big)\Big|_{\text{Ker}J^+}
-\text{Tr}_{\boldsymbol\Xi}\big(e^{t\susy^2}\big)\Big|_{\text{Ker}J^-}
 \nonumber \\ &=&
 \sum_{n\ge0}\left(x^{n+|s|-ia+\rch}-x^{-n-|s|-ia+\rch-1}\right),\quad
 x\equiv e^{\frac{it}\ell}\,.
\end{eqnarray}
This is for the chiral multiplet with unit charge under a $U(1)$ vector
multiplet, and the generalization is straightforward. For chiral
multiplet in a representation $\Lambda$ of the gauge group, the
character at the saddle point $(a,s)$ is given by the sum over the
weight vectors $w\in\Lambda$,
\begin{eqnarray}
 \chi_{\Lambda,\rch}(a,s) &=&
 \sum_{w\in\Lambda}\sum_{n\ge0}
 \left(x^{n+|s\cdot w|-ia\cdot w+\rch}
      -x^{-n-|s\cdot w|-ia\cdot w+\rch-1}\right).
\end{eqnarray}

\paragraph{Vector multiplet and gauge fixing.}

Let us turn to the vector multiplet. A convenient set of fields is
\begin{equation}
\def\arraystretch{1.1}
\begin{array}{lcl}
 {\bf X}^+ &\equiv& \bar\xi\gamma^m\bar\xi A_m+\bar\xi\gamma^3\bar\xi\rho,\\
 {\bf X}^0 &\equiv& \bar\xi\gamma^m\xi A_m+\bar\xi\gamma^3\xi\rho,\\
 {\bf X}^- &\equiv& \xi\gamma^m\xi A_m+\xi\gamma^3\xi\rho,\\
 {\boldsymbol\Xi} &\equiv&
  \frac12\bar\xi\lambda+\frac12\xi\bar\lambda, \\
 \Sigma &\equiv& \bar\xi\xi\sigma+i\bar\xi\gamma^3\xi\rho-iv^mA_m,
\end{array}
 \quad
\begin{array}{lcl}
 \susy{\bf X}^+ &=& -\tfrac12\bar\xi\bar\lambda,\\
 \susy{\bf X}^0\, &=& \tfrac12\bar\xi\lambda-\tfrac12\xi\bar\lambda,\\
 \susy{\bf X}^- &=& \tfrac12\xi\lambda,\\
 \susy{\boldsymbol\Xi} &=& D+\frac1f\sigma+
  i\bar\xi\gamma^3\xi(F_{12}-\frac\rho f)\,,\\
 \susy\Sigma &=& 0.
\end{array}
\def\arraystretch{1}
\label{vecoh}
\end{equation}
This consists of five Grassmann-even and four Grassmann-odd fields, and
the change of variables from $(A_m,\rho,\sigma,\lambda,\bar\lambda,D)$
to these fields is local and invertible. In addition, for the gauge
fixing we need to introduce the ghost fields $c,\bar c,B$ and the BRST
symmetry $\brs$. The BRST charge $\brs$ acts in the standard way on all
the physical fields, namely as $\text{Gauge}(c)$. For ghost fields, we
follow \cite{Pestun:2007rz} and set the transformation rule on the
saddle point $(a,s)$ as
\begin{equation}
\begin{array}{rcl}
 \brs c &=& cc, \\
 \brs\bar c &=& B, \\
 \brs B &=& 0,
\end{array}
\quad
\begin{array}{rcl}
 \susy c &=& \langle\Sigma\rangle-\Sigma, \\
 \susy\bar c &=& 0, \\
 \susy B &=& v^m\partial_m\bar c+[\langle\Sigma\rangle,\bar c]\,.
\quad
(\langle\Sigma\rangle=a/\ell)
\end{array}
\end{equation}
The total supersymmetry $\hat\susy=\susy+\brs$ then satisfies
\begin{equation}
 (\susy+\brs)^2 ~=~ \frac1\ell\left\{
  iJ_3+\frac i2\RV+\text{Gauge}(a) \right\}\,.
\end{equation}
It is important that the constant modes along the direction of Cartan
subalgebra should be excluded from the ghost fields $(c,\bar c,B)$,
since it is the direction along the saddle point locus and should not be
gauge fixed.

The character for the vector multiplet can be most easily evaluated by
taking as the independent variables the three scalar bosons
$({\bf X}^+,{\bf X}^0,{\bf X}^-)$, three scalar fermions
$({\boldsymbol\Xi},c,\bar c)$ and their $\hat\susy$-superpartners.
Notice that ${\bf X}^\mp$ has $\RV=\pm2$
while all other fields have $\RV=0$. The character is therefore the same
as that for the adjoint chiral multiplet with $\rch=1$.

\paragraph{Fixed point formula.}

The character can also be computed by using the Atiyah-Bott fixed point
formula, and this technique can be applied also to the theory on
squashed spheres. In this way one can also show that the partition
function does not depend on the detail of the squashing as long as
it preserves smoothness. To understand
this technique, recall that the character is the difference
of traces, where the operator of interest $e^{t\susy^2}$ involves a
finite rotation $\varphi\to\varphi+t/\ell$. The trace of such an
operator can be expressed as a sum of contributions from the fixed
points, the north and the south poles. Let
$z\equiv\tan\frac\theta2e^{i\varphi}$ be the local complex coordinate
near the north pole, which is transformed by
$e^{t\susy^2}$ as
\begin{equation}
  z\to\tilde z= zx.\quad (x\equiv e^{it/\ell})
\end{equation}
The north pole contribution to the trace (= diagonal sum) of the
operator $e^{t\susy^2}$ is then given by the integral of
$\delta^2(z-\tilde z)$ multiplied by its value at the north pole.
It should be evaluated in a gauge in which there is no Dirac string
singularity at the north pole, which is related to (\ref{u1gauge}) by
the gauge transformation $e^{-is\varphi}$. The differential operators
$J^a_{[\north]} \equiv e^{-is\varphi}J^a e^{is\varphi}$ in this gauge
behave near the north pole as,
\begin{equation}
 J^+_{[\north]} \simeq \frac\partial{\partial\bar z},\quad
 J^-_{[\north]} \simeq -\frac\partial{\partial z},\quad
 J^3_{[\north]} = z\frac\partial{\partial z}
 -\bar z\frac\partial{\partial\bar z}+s\,.
\end{equation}
The north pole contribution to the character (for the chiral multiplet
with unit $U(1)$ charge) is then
\begin{equation}
\chi\big|_\north ~=~
 \left(x^{s-ia+\rch}-x^{s-ia+\rch-1}\right)\cdot \int
 d^2z\delta^2(z-xz)
~=~
\frac{x^{s-ia+\rch}}{1-x}.
\end{equation}
Note that, depending on how to expand into series in $x$, this result
can be interpreted as either the contribution of the $J^+$-zeromodes of
${\bf X}$ or the $J^-$-zeromodes of ${\boldsymbol\Xi}$. The south pole
contribution can be evaluated in the same way: the differential operators
$J^a_{[\south]}\equiv e^{is\varphi}J^a e^{-is\varphi}$ behave around
there as
\begin{equation}
 J^+_{[\south]} \simeq -\frac\partial{\partial w},\quad
 J^-_{[\south]} \simeq  \frac\partial{\partial\bar w},\quad
 J^3_{[\south]} = -w\frac\partial{\partial w}
 +\bar w\frac\partial{\partial\bar w}-s\quad
 (w\equiv\cot\tfrac\theta2 e^{-i\varphi}).
\end{equation}
which leads to
\begin{equation}
\chi\big|_\south~=~
\frac{x^{-s-ia+\rch}}{1-x}\,.
\end{equation}
Combining the two pole contributions expanded into opposite series in
$x$ we obtain the character,
\begin{equation}
\chi~=~
\sum_{n\ge0}\left(x^{n+s-ia+\rch}-x^{-1-n-s-ia+\rch}\right).
\label{chi-u1}
\end{equation}
It is invariant under $s\to -s$, since $s$ is half-integer valued.
Note that each term in the infinite sum corresponds to an approximate
zeromode of $J^+$ or $J^-$, namely a local holomorphic function,
\begin{equation}
 {\bf X}=z^n,\quad {\boldsymbol\Xi}=w^n,\quad n\in\mathbb Z_{\ge0}
\end{equation}
near one of the poles. Note also that the derivation of the character
does not depend on the detailed form of the squashed metric, as long as
it preserves the $U(1)$ isometry and is regular at the poles.

\paragraph{One-loop determinant and an anomaly.}

The one-loop determinant is evaluated as the product of the eigenvalues
of $\susy^2$. For the unit $U(1)$-charge chiral multiplet, the character
(\ref{chi-u1}) leads to the determinant
\begin{equation}
 Z_\text{1-loop}~=~ \prod_{n\ge0}\frac{-n-1-s-ia+\rch}{n+s-ia+\rch}
 ~=~ (\text{phase})\cdot \frac{\Gamma(s-ia+\rch)}{\Gamma(s+ia+1-\rch)},
\end{equation}
where we sign-flipped all the eigenvalues in the enumerator of the
left hand side of the second equality. To keep the invariance of the
determinant under $s\to -s$, one needs an $s$-dependent phase on the
right hand side. However, this symmetry is actually anomalous in the
path integral, and the formula
\begin{equation}
 Z_\text{1-loop}(s,a,\rch)
 ~=~\frac{\Gamma(s-ia+\rch)}{\Gamma(s+ia+1-\rch)}
 ~=~Z_\text{1-loop}(-s,a,\rch)(-1)^{2s},
\label{ccanom}
\end{equation}
gives the correct one-loop determinant which leads to the nice
factorization property of sphere partition function
\cite{Benini:2012ui,Doroud:2012xw}.

To elaborate on this anomaly, let us look into the path integral with
respect to the fermion in the chiral multiplet, which is
\begin{equation}
\int D\bar\psi D\psi \exp\left(-\int d^2x\sqrt g{\cal L}\right),
\quad
{\cal L}\equiv
-\bar\psi\left\{\gamma^mD_m-i\gamma^3\rho-\sigma
-\frac{i\rch}{\ell}\right\}\psi,
\end{equation}
with the vector multiplet fields fixed at the saddle point value
(\ref{u1gauge}). We notice that the squashed sphere geometry (\ref{BG1})
is invariant under the antipodal map
\[
 (\theta,\varphi)\to(\pi-\theta,\pi+\varphi),
\]
if $f(\theta)=f(\pi-\theta)$. The Lagrangian ${\cal L}$ is also
invariant if the antipodal map is defined to act on fields of various
spins as,
\begin{equation}
\begin{array}{ccrcl}
 \text{scalar} &:&
 \tilde\phi(\theta,\varphi) &=& \phi(\pi-\theta,\varphi+\pi),\\
 \text{spinor} &:&
 \tilde\psi(\theta,\varphi) &=&
  i\gamma^2\psi(\pi-\theta,\varphi+\pi),\\
 \text{vector} &:&~~~
 (\tilde A_\theta,\tilde A_\varphi)(\theta,\varphi) &=&
 (-A_\theta,A_\varphi)(\pi-\theta,\varphi+\pi).
\end{array}
\label{antip}
\end{equation}
One can easily check that the Killing spinors (\ref{KS}) are invariant
under this map, and the saddle point configuration (\ref{saddle})
labeled by $(s,a)$ is transformed to another saddle point with $s$
sign-flipped.  This symmetry would lead to the invariance of
$Z_\text{1-loop}$ under $s\to -s$ if there were no anomaly. The sign
factor $(-1)^{2s}$ should arise from the non-invariance of the measure.

To see the anomaly explicitly, let us expand the fermions
$\psi=(\psi^+,\psi^-)$ into the eigenfunctions of chiral Dirac operator
$D_\pm^{(s)}$ squared,
\begin{equation}
 \psi^\pm(\theta,\varphi)
 = \sum_i \psi^\pm_i Y^{s,\pm}_i(\theta,\varphi),\quad
 D^{(s)}_{\pm}D^{(s)}_{\mp}Y^{s,\pm}_i = -\lambda_i Y^{s,\pm}_i.
\label{modeexp}
\end{equation}
Let $n_\pm$ be the numbers of zeromodes for the chiral (antichiral)
components $\psi^\pm$. The integration measure for $\psi$ then takes the
form
\begin{eqnarray}
 D\psi \equiv
 \prod_{\lambda_i=0}d\psi^+_i\cdot
 \prod_{\lambda_i=0}d\psi^-_i\cdot
 \prod_{\lambda_i\ne0}d\psi_i^+d\psi_i^-\,,
\end{eqnarray}
where the three factors correspond to $n_+$ zeromodes of $\psi^+$,
$n_-$ zeromodes of $\psi^-$ and the rest (nonzero modes). Note that
$n_+-n_-=-2s$ from the index theorem. The fermions $\bar\psi^\alpha$ are
expanded into the eigenfunctions $Y^{-s,\pm}_i$ in the same way.

The eigenfunctions $Y^{s,+}_i$, $Y^{s,-}_i$ of the same nonzero
eigenvalue $\lambda_i$ are paired by the action of $D_\pm^{(s)}$.
They obey the periodicity
\begin{equation}
 Y^{s,\pm}_i(\theta,\varphi+2\pi)=
 (-1)^{2s+1}Y^{s,\pm}_i(\theta,\varphi)\,,
\end{equation}
if the background spin connection and the vector multiplet fields are in
the gauge (\ref{BG1}), (\ref{saddle}).
Also, the antipodal map acts on the chiral Dirac operators as
$D_\pm^{(s)}\to-D_\mp^{(-s)}$. Combining all these one can show that the
set of eigenfunctions can be chosen to satisfy
\begin{equation}
\begin{array}{lcrcr}
 2s=\text{even}&:~~~& Y^{s,\pm}_i(\pi-\theta,\pi+\varphi)
 &=& i Y^{-s,\mp}_i(\theta,\varphi),\\
 2s=\text{odd}&:~~~& Y^{s,\pm}_i(\pi-\theta,\pi+\varphi)
 &=&   Y^{-s,\mp}_i(\theta,\varphi).
\end{array}
\end{equation}

The antipodal map transforms $\psi$ to $\tilde\psi$ as in (\ref{antip}).
Their modes $\tilde\psi_i^\pm$ are defined and related to the modes
of the original fermion $\psi_i^\pm$ as follows,
\begin{equation}
  \tilde\psi^\pm(\theta,\varphi)
 =\sum_i\tilde\psi_i^\pm Y_i^{-s,\pm}(\theta,\varphi),\quad
 \tilde\psi_i^\pm=
 \left\{\begin{array}{ll}
 \pm i\psi_i^\mp ~~&(2s=\text{even}) \\
 \pm \psi_i^\mp &(2s=\text{odd})
	\end{array}\right.\,.
\end{equation}
This leads to the anomaly
\begin{equation}
 D\tilde{\bar\psi} D\tilde\psi = (-1)^{2s} D\bar\psi D\psi
\end{equation}
as claimed, where the sign factor arises from the zeromode part of the measure.
Let us note here that one can work out the full eigenfunctions and
eigenmodes explicitly for the round sphere, since in that case the
operators $D_\pm^{(s)}$ coincides with those defined in (\ref{Dirac})
and the eigenfunctions $Y^{s,\pm}_i$ are nothing but the monopole
harmonics $Y^{s\pm1/2}_{jm}$. The monopole harmonics with $j=|s|-1/2$
are the fermion zeromodes.

\paragraph{Summary.}

The one-loop determinant for general chiral multiplet of the R-charge
$\RV=2\rch$ in the representation $\Lambda$ of the gauge group is,
\begin{equation}
 Z_{\text{ch},\Lambda}(s,a,\rch)
 = \prod_{w\in\Lambda}Z_\text{1-loop}(s\sdot w\,,\,a\sdot w\,,\rch)
 = \prod_{w\in\Lambda}
 \frac{\Gamma(s\sdot w-ia\sdot w+\rch)}
      {\Gamma(s\sdot w+ia\sdot w+1-\rch)},
\end{equation}
where the saddle-point parameters $a,s$ are Cartan subalgebra valued,
and $w$ runs over the weight vectors of the representation $\Lambda$.
For the vector multiplets for simple Lie algebras the determinant is
\begin{equation}
 Z_\text{vec}(s,a)=Z_{\text{ch, adj}}(s,a,\rch=1)
 = \prod_{\alpha\in\Delta_+}\left\{(a\sdot\alpha)^2+(s\sdot\alpha)^2\right\}\,,
\end{equation}
where the product runs over positive roots. The Coulomb branch
localization formula for the partition function thus becomes
\begin{equation}
 Z_{S^2}~=~ \frac1{|W|}\sum_s\int\frac{d^ra}{(2\pi)^r}
 z^{\text{Tr}(ia-s)}\bar z^{\text{Tr}(ia+s)}
 Z_\text{vec}(s,a)Z_{\text{ch},\Lambda}(s,a,\rch),
\end{equation}
where $z=\exp(-2\pi\zeta-i\theta)$ and $r$ is the rank of the gauge
group. For theories with flavor symmetries, one can introduce matter
masses by gauging it by an external vector multiplet and turning
on its $\sigma$-components. The matter one-loop determinant will then
depend on the masses in an obvious manner.

Let us note here that, due to the anomaly in the antipodal map
(\ref{ccanom}), the partition function is not invariant under a simple
sign-flip of $\theta$ angle or $z\leftrightarrow\bar z$. For example,
for a $U(1)$ gauge theory with $N_\text{F}$ electrons (chiral multiplets with
charge $+1$) and $N_\text{A}$ positrons (charge $-1$), the one-loop
determinant from the matters obeys
\begin{equation}
 Z_\text{1-loop}(s,a)=(-1)^{2s(N_\text{F}+N_\text{A})}Z_\text{1-loop}(-s,a),
\end{equation}
so the partition function is invariant under $z\to (-1)^{N_\text{F}+N_\text{A}}\bar z$.

\section{Defects}\label{sec:defect}

In this section we wish to generalize the exact results for partition
function to include the codimension two vortex defects. To keep the
supersymmetry (\ref{KS}) unbroken, we only consider the defects inserted
at the north and the south poles of the squashed sphere. The defects of
interest are characterized by the singular behavior of the gauge field,
\begin{equation}
 A \simeq \eta^\north\cdot d\varphi~~(\theta=0),\qquad
 A \simeq \eta^\south\cdot d\varphi~~(\theta=\pi),
\end{equation}
where the constants $\eta^\north,\eta^\south$ are in Cartan
subalgebra. In local Cartesian coordinates $x^1,x^2$ centered at the
poles we have $F_{12}\simeq 2\pi\eta^\north\delta^2(x)$ or
$-2\pi\eta^\south\delta^2(x)$.
Noticing that our Killing spinors $\xi,\bar\xi$ are purely chiral or
anti-chiral at the poles, the unbroken supersymmetry requires $D$ to
take singular values,
\begin{equation}
 D\simeq 2\pi i\eta^\north\delta^2(x)~~(\theta=0)\;,
 \qquad
 D\simeq 2\pi i\eta^\south\delta^2(x)~~(\theta=\pi)\;.
\end{equation}

\paragraph{Abelian theories.}

Let us begin with the simplest case of the $U(1)$ gauge theory. Up to
gauge equivalence, the saddle point configurations in the presence of
the vortex defects are given by
\begin{eqnarray}
 && \sigma = \frac a\ell,\hskip5.9mm
 D= \frac{a}{f\ell}+2\pi i\eta^\north\delta^2_{(\text{NP})}
    +2\pi i\eta^\south\delta^2_{(\text{SP})},
 \nonumber \\ &&
 \rho=-\frac s\ell,\quad
 A=\left\{\begin{array}{ll}
    s(\cos\theta-1)d\varphi+\eta^\north d\varphi\,, ~~ &
    \text{(north patch)} \\
    s(\cos\theta+1)d\varphi+\eta^\south d\varphi\,. &
    \text{(south patch)}
	  \end{array}\right.
\end{eqnarray}
The quantization condition of magnetic flux gives
$\eta^\north-\eta^\south-2s\in\mathbb Z$.
Using Coulomb branch localization, the correlators of the vortex defects
are thus expressed as
\begin{equation}
 \left\langle
 V_{\eta^\north}
 V_{\eta^\south}\right\rangle =
 \sum_{s\in\frac12(\eta^\north-\eta^\south+\mathbb Z)}\int da
 \;z^{\eta^\north+ia-s}\,\bar z^{\eta^\south+ia+s}\,
 Z_\text{1-loop}\,,
\label{VV}
\end{equation}
For the theories with $U(1)$ gauge group, non-trivial one-loop
determinant arises only from charged chiral multiplets. We would like to
evaluate it by generalizing the analysis of the previous section.

The square of the supersymmetry on this saddle point is
\begin{equation}
 \susy^2=\frac1\ell\left\{
 \partial_\varphi+\frac i2\RV+\text{Gauge}(\tilde a)\right\},\quad
 \tilde a=\left\{\begin{array}{ll}
	   a+is-i\eta^\north & \text{(north patch)} \\
	   a-is-i\eta^\south & \text{(south patch)}
		 \end{array}
 \right.
\end{equation}
By a naive application of the fixed point formula, the contribution of a
chiral multiplet with unit charge and $\RV=2\rch$ to the determinant
would be obtained from the character,
\begin{equation}
 \chi ~=~ \frac{x^{s-\eta^\north-ia+\rch}}{1-x}
+\frac{x^{-s-\eta^\south-ia+\rch}}{1-x}
 ~=~
 \sum_{n\ge0}\left(
x^{n+s-\eta^\north-ia+\rch}-x^{-n-s-\eta^\south-ia+\rch-1}
 \right)\,.
\label{chidfct}
\end{equation}
In the second equality we expanded the first term in the left hand
side into positive series in $x$ and the second term into the negative series.
The other way of expansion gives the same answer thanks to the flux
quantization condition. Assuming this is the correct character, one can
translate it into the determinant
\begin{equation}
 Z_\text{1-loop}=
 \frac{\Gamma(s-\eta^\north-ia+\rch)}
      {\Gamma(s+\eta^\south+ia-\rch+1)}\,.
\end{equation}
This would mean that the effect of vortex defects in one-loop
determinants is simply to shift the
saddle-point parameters. Similar results were obtained for the
expectation values of vortex loop observables in 3D SUSY gauge theories
in \cite{Kapustin:2012iw,Drukker:2012sr}.

A problem with this naive result is that it does not respect the
fact that $\eta^\north,\eta^\south$ are periodic parameters with
period 1. Namely, under the assumption that the $U(1)$ charges of all
the matters are integers, one can shift $\eta^\north,\eta^\south$
by integers by singular gauge transformations without introducing
multi-valuedness to the fields. Also, the normalizability of the modes
(such as $Y_i^{s,\pm}$ in (\ref{modeexp})) would not change under such
gauge transformations, so the determinants should be invariant.
The above naive formula should therefore be at most correct only for
$\eta^\north,\eta^\south$ within a certain range.

In the following we will find the fully general formula with the correct
periodicity in $\eta$'s, by making use of the correspondence between the
gauge theory in the presence of codimension two defects (ramified
bundle) and the gauge theory on orbifolds.

\paragraph{Comparison with SUSY orbifolds.}

Let us consider the special case
\begin{equation}
\eta^\north=\eta^\south=\eta=r/K~(r,K\in\mathbb Z)\,. 
\end{equation}
In this case,
the singularity of the gauge field at the two poles can be removed by
the gauge transformation $A'=A-\eta d\varphi$. But this makes the charged
matter fields non-periodic around the poles. For example, the matter field
$\Phi(\theta,\varphi)$ with unit $U(1)$ charge obeys twisted periodicity
condition,
\begin{equation}
 \Phi(\theta,\varphi+2\pi)=\Phi(\theta,\varphi)e^{-2\pi i\eta}\,.
\end{equation}
Recall that the terms in the character (\ref{chi-u1}) obtained in the
previous section were associated with the basis of local holomorphic
functions at the two poles. The $\eta$-dependence of the character
(\ref{chidfct}) on the defect background can be understood to arise
because the basis holomorphic functions now have to obey twisted
periodicity condition.

We recall that similar twist in periodicity
appears in gauge theories on the orbifold $S^2/\mathbb Z_K$, where the
$\mathbb Z_K$ symmetry act on charged fields as a gauge rotation as
well as the rotation of the sphere
\begin{equation}
 \Phi(\theta,\varphi+\tfrac{2\pi}K)=\Phi(\theta,\varphi)e^{-\frac{2\pi i r}K}\,.
\end{equation}
The orbifolds of sphere can be regarded as squashed spheres with
conical singularities at the two poles. Our idea is thus to compute the
correlators of vortex defects (\ref{VV}) using standard orbifold technique,
and then generalize the result to the spheres with arbitrary
conical singularities at the poles.

We are interested in the orbifold of $S^2$ preserving supersymmetry. Our
Killing spinors $\xi,\bar\xi$ (\ref{KS}) are not invariant under the
$\mathbb Z_K$ rotation $\varphi\to\varphi+\frac{2\pi}K$, but they satisfy
\begin{equation}
 \xi(\theta,\varphi+\tfrac{2\pi}K)=e^{-\frac{i\pi}K}\xi(\theta,\varphi),\quad
 \bar\xi(\theta,\varphi+\tfrac{2\pi}K)=e^{\frac{i\pi}K}\bar\xi(\theta,\varphi).
\end{equation}
So the supersymmetric $\mathbb Z_K$ orbifold should involve the vector
R-symmetry rotation. At first sight it seems natural to impose the
orbifold projection
\begin{equation}
 \Phi(\theta,\varphi+\tfrac{2\pi}K) \stackrel{?}=
 e^{-\frac{i\pi}K\cdot\RV-\frac{2\pi ir}K\cdot Q}\,\Phi(\theta,\varphi)
\end{equation}
on the field $\Phi$ of electric charge $Q$. However, this condition is
not consistent with the periodicity of the fields on the sphere before
orbifolding,
\begin{equation}
 \Phi(\theta,\varphi+2\pi)=(-1)^{2\times\text{(spin)}}\,\Phi(\theta,\varphi),
\end{equation}
where the sign factor arises because of our choice of local Lorentz
frame in which the spin connection takes the form (\ref{BG1}). It turns
out that the SUSY-preserving orbifold which is suitable for our purpose is
\begin{equation}
 \Phi(\theta,\varphi+\tfrac{2\pi}K) = (-1)^{2\times\text{(spin)}}\,
 e^{\frac{i\pi(K-1)}K\cdot\RV-\frac{2\pi ir}K\cdot Q}\,\Phi(\theta,\varphi).
\label{orb}
\end{equation}
In order for this to be a $\mathbb Z_K$ orbifold projection, one also
needs to impose that $K$ is odd and the label $\rch$ of all the chiral
multiplets has to satisfy $(K-1)\rch\in\mathbb Z$. But even if these
conditions are not met, the condition (\ref{orb}) can be interpreted as
the periodicity condition for the fields on a squashed sphere,
\begin{equation}
 ds^2 = f^2 d\theta^2 + \ell^2\sin^2\theta d\varphi^2,
 \quad
 V=\frac12\left(\frac \ell f-1\right)d\varphi,\quad
(f\equiv K\ell)
\label{consq}
\end{equation}
with vortex defects at the two poles. The $\RV$-charge dependence of the
projection condition (\ref{orb}) arises after one gauges away the
background field $V$. This interpretation works for arbitrary real
positive $K$.

Let us work out the orbifolded character for a chiral multiplet with
unit electric charge and $\RV=2\rch$. The projection condition on the
variables ${\bf X},\boldsymbol\Xi$ becomes
\begin{eqnarray}
 {\bf X}(\theta,\varphi+\tfrac{2\pi}K) &=&
 e^{\frac{2\pi i}K(K-1)\rch-\frac{2\pi i}Kr}{\bf X}(\theta,\varphi),
 \nonumber \\
 {\boldsymbol\Xi}(\theta,\varphi+\tfrac{2\pi}K) &=&
 e^{\frac{2\pi i}K(K-1)(\rch-1)-\frac{2\pi i}Kr}
 {\boldsymbol\Xi}(\theta,\varphi).
\end{eqnarray}
We assume $(K-1)\rch$ is an integer, and also that the R-charge takes
a reasonable value $0\le2\rch\le2$. The orbifolded character is then
obtained by projecting to the $\mathbb Z_K$-invariant modes,
\begin{equation}
 \chi ~=~
 \sum_{n\ge0}^\spadesuit x^{n+s-ia+\rch}
-\sum_{n\ge 0}^\heartsuit x^{-n-s-ia+\rch-1}\,. \quad
\left(\begin{array}{cl}
 \mathsmaller{\spadesuit~:} & \mathsmaller{n=\rch(K-1)-r\text{ mod }K} \\
 \mathsmaller{\heartsuit~:} & \mathsmaller{n=-(\rch-1)(K-1)+r\text{ mod }K}
       \end{array}\right)
\end{equation}
By introducing the rescaled variables
\[
 x^K=\tilde x,\quad
 \frac rK=\eta,\quad
 \frac sK=\tilde s,\quad
 \frac aK=\tilde a,
\]
one can rewrite the character as follows,
\begin{equation}
 \chi ~=~
 \sum_{n\ge \eta-\rch(1-\frac1K),\;n\in\mathbb Z}
 \tilde x^{n+\tilde s-\eta-i\tilde a+\rch}
-\sum_{n\le \eta+(1-\rch)(1-\frac1K),\;n\in\mathbb Z}
 \tilde x^{n-\tilde s-\eta-i\tilde a+\rch-1}\,.
\label{ch-orb1}
\end{equation}
The rescaling of $x$ is to adjust the radius of the orbifold $S^2/\mathbb Z_K$
to that of the squashed sphere (\ref{consq}). Note also that $s$ is the
magnetic flux through the $S^2$, whereas the flux through the orbifold is
$\tilde s$ which is to be GNO quantized.

Now we would like to generalize this result to arbitrary real positive $K$. The orbifolded character (\ref{ch-orb1}) consists of two infinite sums, each depending on $K$ through the restriction on the integer $n$. If we generalize this formula to non-integer $K$ as it is, the result will not be invariant under $s\to -s$. Using $\lfloor x+1-\frac1K\rfloor=\lceil x\rceil$ for $x\in\frac1K\mathbb Z$, we notice the formula (\ref{ch-orb1}) can be rewritten in such a way that the invariance under $s\to -s$ is maintained,
\begin{equation}
\chi = \sum_{n\ge [\eta]_\rch,\,n\in\mathbb Z}
\tilde x^{n+\tilde s-\eta-i\tilde a+\rch}
 -\sum_{n\le[\eta_\rch],\, n\in\mathbb Z}
\tilde x^{n-\tilde s-\eta-i\tilde a+\rch-1},
\label{ch-orb2}
\end{equation}
where we have two possible definitions of the integer-valued function
$[\eta]_\rch$,
\begin{equation}
 ({\rm I})\quad
 [\eta]_\rch\equiv\lceil\eta-\rch(1-\tfrac1K)\rceil\quad\text{or}\quad
 ({\rm II})\quad
 [\eta]_\rch\equiv\lfloor\eta+(1-q)(1-\tfrac1K)\rfloor.
\label{norf}
\end{equation}
These two choices indicate there are two possible boundary conditions for the matter fields near vortex defects. In \cite{Hosomichi:2017dbc} these boundary conditions were named ``normal'' and ``flipped'', respectively. For more general defect backgrounds with $\eta^\north\ne\eta^\south$, we propose that the character and the corresponding one-loop determinant are given by
\begin{eqnarray}
 && \chi = \sum_{n\ge0}
 \left(x^{ n+s-\eta^\north-ia+\rch+[\eta^\north]_{\rch}}
      -x^{-n-s-\eta^\south-ia+\rch-1+[\eta^\south]_{\rch}}\right),
\nonumber \\
 && Z_\text{1-loop} =
 \frac{\Gamma(s-\eta^\north-ia+\rch+[\eta^\north]_{\rch})}
      {\Gamma(1+s+\eta^\south+ia-\rch-[\eta^\south]_{\rch})}.
\label{Z1lwv}
\end{eqnarray}
Here $[\eta^\north]_{\rch}, [\eta^\south]_{\rch}$ are defined as (\ref{norf}). They depend on the vorticity and the matter $\RV$-charge $\rch$ as well as the deficit angle $K$ and the choice of boundary condition at the poles. Note also that the character has the expected periodicity in $\eta$.

For general $K$ the spectrum of wavefunctions does not follow from orbifold projection, and actually the index formula (\ref{ch-orb2}) and (\ref{Z1lwv}) indicate that some of the modes contributing to the index are divergent near the poles. To be more explicit, the local (anti-)holomorphic functions near the north pole that contribute to the index are
\begin{equation}
 {\bf X} = z^{n+[\eta^\north]_{\rch}-\eta^\north},\quad
 {\boldsymbol\Xi} = \bar z^{n-[\eta^\north]_{\rch}+\eta^\north}\quad
 (n\in\mathbb Z_{\ge0}).
\end{equation}
More specifically, for $K=1$ the zeromodes of $J^\pm$ that contribute to the index are given by
\begin{alignat}{4}
 \text{(I) normal b.c.}&:&\quad
 {\bf X} &=z^{n+\lceil\eta^\north\rceil-\eta^\north},&\quad
 {\boldsymbol\Xi} &=\bar z^{n-\lceil\eta^\north\rceil+\eta^\north}, \nonumber\\
 \text{(II) flipped b.c.}&:&\quad
 {\bf X} &=z^{n+\lfloor\eta^\north\rfloor-\eta^\north},&\quad
 {\boldsymbol\Xi} &=\bar z^{n-\lfloor\eta^\north\rfloor+\eta^\north}.&\quad
 (n\in\mathbb Z_{\ge0})
\end{alignat}
Namely, the normal boundary condition requires ${\bf X}$ to be finite at $z=0$ while ${\boldsymbol\Xi}$ is allowed to diverge mildly as ${\boldsymbol\Xi}\sim\bar z^\gamma\,(\gamma>-1)$. Similarly, the flipped boundary condition requires ${\boldsymbol\Xi}$ to be finite while ${\bf X}$ is allowed to diverge mildly.

\paragraph{Example 1: SQED.}

As an application, let us study the vortex defects in the
SQED with $N_\text{F}$ electrons and $N_\text{A}$ positrons. To make the
formulae short, we combine the masses $\mu$ and the
$\RV$-charges $\rch$ of the matters into complex parameters $m$,
\begin{equation}
 m_i\equiv \mu_i + i\rch_i~~(i=1,\cdots, N_\text{F}),\qquad
 \tilde m_i\equiv \tilde\mu_i + i\tilde\rch_i~~(i=1,\cdots, N_\text{A}).
\end{equation}
The correlator of the vortex defects can be expressed as
\begin{eqnarray}
 \langle V_{\eta^\north}V_{\eta^\south}\rangle
 &=& \sum_{s\in\frac12(\eta^\north-\eta^\south+\mathbb Z)}
 \int\frac{da}{2\pi}\,
 z^{-s+\eta^\north+ia}\bar z^{s+\eta^\south+ia}
 \cdot Z^{\text{\smaller SQED}}_\text{ch},
 \nonumber \\
 Z^{\text{\smaller SQED}}_\text{ch} &\equiv& 
 \prod_{i=1}^{N_\text{F}}
 \frac{\Gamma(s-\eta^\north-ia-im_i+[\eta^\north]_{\rch_i})}
      {\Gamma(1+s+\eta^\south+ia+im_i-[\eta^\south]_{\rch_i})}
 \nonumber \\ && \hskip-2.1mm\cdot
 \prod_{i=1}^{N_\text{A}}
 \frac{\Gamma(-s+\eta^\north+ia-i\tilde m_i
              +[-\eta^\north]_{\tilde\rch_i})}
      {\Gamma(1-s-\eta^\south-ia+i\tilde m_i
              -[-\eta^\south]_{\tilde\rch_i})}.
\end{eqnarray}
There are two questions we would like to address. One is whether the apparent
dependence on the $\eta$'s can be eliminated by a redefinition of
contours, or there is nontrivial $\eta$-dependence remaining. The other
is whether the Higgs branch expression has a factorized form as in the
absence of the defects.

The contour of $a$-integration can be closed in the lower or upper
half-planes depending on whether $|z|<1$ or $|z|>1$
\cite{Benini:2012ui,Doroud:2012xw}. Let us focus on the former case. The
contour integral then picks up the poles of the determinants of
$N_\text{F}$ electrons. The determinant of the $j$-th electron has a set
of poles labeled by a pair of non-negative integers
$(k_\north,k_\south)$. The saddle point parameters $a$ and $s$ are
related to them by
\begin{eqnarray}
 -s+\eta^\north+ia &=& -im_j+[\eta^\north]_{\rch_j}+k_\north,
  \nonumber \\
  s+\eta^\south+ia &=& -im_j+[\eta^\south]_{\rch_j}+k_\south.
\end{eqnarray}
One can check the value of $a$ is always in the lower half plane as long as $q_j>0$ provided the electron obeys the normal boundary condition. Let us choose $q_j,\tilde q_j$ and the boundary conditions so that the poles of the determinant from electrons are all in the lower half-plane and those from positrons are all in the upper half-plane.

The sum over residues can then be organized into the following form
\begin{eqnarray}
 \langle V_{\eta^\north}V_{\eta^\south}\rangle
 &=& \sum_{j=1}^{N_\text{F}}(-1)^{[\eta^\north]_{\rch_j}}
 \prod_{i\ne j}^{N_\text{F}}(-1)^{[\eta^\north]_{\rch_i}}
 \frac{\Gamma(-im_i+im_j)}{\Gamma(1+im_i-im_j)}
 \prod_{i=1}^{N_\text{A}}(-1)^{[-\eta^\south]_{\tilde\rch_i}}
 \frac{\Gamma(-i\tilde m_i-im_j)}{\Gamma(1+i\tilde m_i+im_j)}
\nonumber \\ && \times |z|^{-2im_j}
 F^{[j]}_\text{vortex}
 \left((-1)^{N_\text{F}}z\,;\,\eta^\north,m_i,\tilde m_i \right)
 F^{[j]}_\text{vortex}
 \left((-1)^{N_\text{A}}\bar z\,;\,\eta^\south,m_i,\tilde m_i \right),
\end{eqnarray}
where $F^{[j]}_\text{vortex}$ is a generalization of the vortex
partition function for the SQED,
\begin{eqnarray}
 F^{[j]}_\text{vortex}
 \left(z\,;\,\eta,m_i,\tilde m_i \right)
 &=& 
 \sum_{k\ge [\eta]_{q_j}}\!
 \frac{z^k}{(k-[\eta]_{q_j})!}
 \frac{\prod_{i=1}^{N_\text{A}}
       (-i\tilde m_i-im_j)_{k+[-\eta]_{\tilde\rch_i}}}
      {\prod_{i\ne j}^{N_\text{F}}
       (1+im_i-im_j)_{k-[\eta]_{\rch_i}}},
\label{Fv}
\end{eqnarray}
and $(x)_n=\Gamma(x+n)/\Gamma(x)$ is the Pochhammer symbol (we allow the
integer $n$ to become negative).
Note that, in the product of two vortex partition functions, one
depends only on $\eta^\north$ while the other depends only on
$\eta^\south$. Therefore each vortex partition function
captures the physics near one of the poles as expected. Note also that
\begin{equation}
 F^{[j]}_\text{vortex}(z;\eta+1,m_i,\tilde m_i)
 = z\cdot F^{[j]}_\text{vortex}(z;\eta,m_i,\tilde m_i),
\end{equation}
thanks to the periodicity of the matter path integral in $\eta$.

If there is no conical singularity $(K=1)$, the integer-valued function $[\eta]_{\rch_i}$ all become $\rch_i$-independent and $F^{[j]}_\text{vortex}$ simplifies significantly. For example, if all the matters obey normal boundary condition, we have
\begin{equation}
 F^{[j]}_\text{vortex} = z^{\lceil\eta\rceil}\sum_{k\ge0}\frac{z^k}{k!}
\frac{\prod_{i=1}^{N_{\text{A}}}(-i\tilde m_i-im_j)_{k+\delta_\eta}}
     {\prod_{i\ne j}^{N_\text{F}}(1+im_i-im_j)_{k}},\quad
 \delta_\eta= \left\{\begin{array}{ll}
 0 & (\eta\in\mathbb Z)	\\ 1 & (\eta\nin\mathbb Z) \end{array}\right. .
\end{equation}

\paragraph{Non-Abelian theories.}

Let us next turn to non-Abelian gauge theories. The saddle point configurations are parametrized by ${\boldsymbol s}, {\boldsymbol a}$ and the vorticity parameters ${\boldsymbol\eta}^\north,{\boldsymbol\eta}^\south$, all taking values in Cartan subalgebra. An important subtlety here is that summing over different saddle points involves summation over Weyl images of ${\boldsymbol\eta}^\north$ and ${\boldsymbol\eta}^\south$, though any two saddle points related to each other by a simultaneous Weyl reflections of ${\boldsymbol a}, {\boldsymbol s}, {\boldsymbol\eta}^\north, {\boldsymbol\eta}^\south$ are of course equivalent.

The character and one-loop determinant for the fluctuations of fields around saddle points can be obtained by generalizing the result for the abelian case. For a chiral multiplet in the representation $\Lambda$ and $\RV=\rch$, the character is given by
\begin{equation}
 \chi_{_{\Lambda,\rch}} =
 \sum_{w\in\Lambda}\sum_{n\ge0}\left(
 x^{n+q+w\cdot({\boldsymbol s}-{\boldsymbol\eta}^\north-i{\boldsymbol a})
    +[w\cdot{\boldsymbol\eta}^\north]_q}
-x^{-n+q-1-w\cdot({\boldsymbol s}+{\boldsymbol\eta}^\south+i{\boldsymbol a})
    +[w\cdot{\boldsymbol\eta}^\south]_q}
\right),
\end{equation}
with $w$ the weight vectors, and the determinant is
\begin{equation}
 Z_{\text{ch},\Lambda} =
 \prod_{w\in\Lambda}
 \frac
 {\Gamma(q+w\!\cdot\!
 ({\boldsymbol s}-{\boldsymbol\eta}^\north-i{\boldsymbol a})
 +[w\!\cdot\!{\boldsymbol\eta}^\north]_q)}
 {\Gamma(1-q+w\!\cdot\!
 ({\boldsymbol s}+{\boldsymbol\eta}^\south+i{\boldsymbol a})
 -[w\!\cdot\!{\boldsymbol\eta}^\south]_q)}.
\end{equation}
For the vector multiplet, one can move to the set of fields introduced at (\ref{vecoh}) and identify the determinant with that of an adjoint chiral multiplet with $\rch=1$. We still need to find out the appropriate boundary condition on the fields. From the behavior of the fields ${\bf X}^a$ near the north pole
\begin{equation}
 {\bf X}^+\simeq \frac1\ell A_{\bar z},\quad
 {\bf X}^-\simeq -\frac1\ell A_z,\quad
 {\bf X}^0\simeq -\frac1\ell A_\varphi-\rho,
\end{equation}
it is reasonable to allow that the fluctuation of ${\bf X}^\pm$ may diverge mildly near the north pole, but the fluctuation of ${\bf X}^0$ should be finite. Also, at saddle points the scalars $\sigma,\rho,D$ take constant values in Cartan subalgebra, which makes sense only if we require their fluctuations to be finite everywhere. These are identified with the behavior for an adjoint chiral multiplet satisfying flipped boundary condition. The determinant then turns out to be $K$-independent,
\begin{eqnarray}
 Z_\text{vec} &=& \prod_{\alpha\in\Delta}
 \frac{\Gamma(1+\alpha\sdot({\boldsymbol s}-{\boldsymbol\eta}^\north-i{\boldsymbol a})+\lfloor\alpha\sdot{\boldsymbol\eta}^\north\rfloor)}
 {\Gamma(\alpha\sdot({\boldsymbol s}+{\boldsymbol\eta}^\south+i{\boldsymbol a})-\lfloor\alpha\sdot{\boldsymbol\eta}^\south\rfloor)}
 \nonumber \\ &=&
 (-1)^{2\rho\sdot(2{\boldsymbol s}+{\boldsymbol\eta}^\south-{\boldsymbol\eta}^\north)}\cdot
\prod_{\alpha\in\Delta_+}(-1)^{\lceil\alpha\sdot{\boldsymbol\eta}^\north\rceil}
\prod_{\alpha\in\Delta_+}(-1)^{\lceil\alpha\sdot{\boldsymbol\eta}^\south\rceil}
\nonumber \\ && \hskip3mm\cdot
 \prod_{\alpha\in\Delta_+,\alpha\sdot{\boldsymbol\eta}^\north\in\mathbb Z}
 \alpha\cdot({\boldsymbol s}-i{\boldsymbol a})
 \prod_{\alpha\in\Delta_+,\alpha\sdot{\boldsymbol\eta}^\south\in\mathbb Z}
 \alpha\cdot({\boldsymbol s}+i{\boldsymbol a}).
\end{eqnarray}
Here $\Delta\,(\Delta_+)$ are the set of (positive) roots and $\rho$ is the Weyl vector. The first factor in the second line is trivial for simple Lie algebras due to quantization of magnetic flux.

As a concrete example, let us concentrate on the $U(N)$ gauge group in the
following. The parameters ${\boldsymbol\eta}^\north$, ${\boldsymbol\eta}^\south$, ${\boldsymbol s}$ and ${\boldsymbol a}$ are now $N\times N$ diagonal matrices,
\def\arraystretch{1.2}
\[
\begin{array}{rcl}
 {\boldsymbol\eta}^\north &=& \text{diag}(\eta^\north_1,\cdots,\eta^\north_N),\\
 {\boldsymbol\eta}^\south &=& \text{diag}(\eta^\south_1,\cdots,\eta^\south_N),
\end{array}
\quad
\begin{array}{rcl}
 \boldsymbol s &=& \text{diag}(s_1,\cdots,s_N), \\
 \boldsymbol a &=& \text{diag}(a_1,\cdots,a_N)\,.
\end{array}
\]
The flux quantization condition reads $\boldsymbol\eta^\north-\boldsymbol\eta^\south-2\boldsymbol s\in\mathbb Z^N$. The defects at the poles break the gauge symmetry to a subgroup called Levi subgroup. Generically $U(N)$ is broken to $U(1)^N$, but some non-Abelian symmetry remains if the diagonal elements of $\boldsymbol\eta^\north$ or $\boldsymbol\eta^\south$ degenerate. For simplicity let us assume $\eta^\north_a,\eta^\south_a$ are all within $[0,1)$. Then the determinant $Z_\text{vec}$ takes the following simple form,
\begin{eqnarray}
 Z_\text{vec} &=& (-1)^{(N-1)\text{Tr}(2{\boldsymbol s}+{\boldsymbol\eta}^\south-{\boldsymbol\eta}^\north)}
 \cdot(-1)^{\sigma({\boldsymbol\eta}^\north)+\sigma({\boldsymbol\eta}^\south)}
 \nonumber \\ && \times
 \prod_{a<b,\eta_{ab}^\north=0}(s_{ab}-ia_{ab})\cdot
 \prod_{a<b,\eta_{ab}^\south=0}(s_{ab}+ia_{ab}),
 \nonumber \\ &&
 \sigma({\boldsymbol\eta})\equiv\{\text{number of pairs $(a,b)$ such that $a<b,\eta_a>\eta_b$}\}.
\label{Zvwd}
\end{eqnarray}
Here we used the short-hand notation $s_{ab}\equiv s_a-s_b$, etc. The first sign factor in the right hand side of (\ref{Zvwd}) can be absorbed into a redefinition of the FI-theta parameter: $z(-1)^{N-1}\to z$.

The discontinuity of the determinant is at
$\eta^\north_{ab}=0$ or $\eta^\south_{ab}=0$, namely when the
eivenvalues of ${\boldsymbol\eta}^{\north,\south}$ degenerate.
The properties of vortex defects are therefore qualitatively
different depending on the choice of Levi subgroups. Let us now take two
vortex defects $V_{\boldsymbol\eta^\north}$,
$V_{\boldsymbol\eta^\south}$ corresponding to the Levi subgroups
$\mathbb L^\north$ and $\mathbb L^\south$, and take the parameters
$\boldsymbol\eta^\north, \boldsymbol\eta^\south$ to be vanishingly
small. In this limit, the one-loop determinant of vector multiplet does
not agree with the one in the absence of the defect. It rather
satisfies
\begin{eqnarray}
 Z_\text{vec}({\mathbb L}^\north,{\mathbb L}^\south;
              \boldsymbol\eta^\north, \boldsymbol\eta^\south)
&\stackrel{\boldsymbol\eta^\north, \boldsymbol\eta^\south\to0}\longrightarrow&
 Z_\text{vec}(\boldsymbol\eta^\north=\boldsymbol\eta^\south=0)
 \cdot (-1)^{\sigma({\boldsymbol\eta}^\north)+\sigma({\boldsymbol\eta}^\south)}
 \nonumber \\ && \cdot
 \prod_{\substack{\alpha\in\Delta_+\\\alpha\nin \mathbb L^\north}}
 \frac1{\alpha\sdot(s-ia)}
 \cdot\!\!\prod_{\substack{\alpha\in\Delta_+\\\alpha\nin \mathbb L^\south}}
 \frac1{\alpha\sdot(s+ia)}\,.
\label{tcr}
\end{eqnarray}
The products are over the positive roots of $U(N)$ which are broken due to the defects. Recall here that $s\mp ia$ are proportional to the saddle point values of $\sigma\mp i\rho$. The relation (\ref{tcr}) therefore implies that the vortex defects have the same effect as inserting certain local operators made of $\sigma-i\rho$ at the north pole, $\sigma+i\rho$ at the south pole. Usually we require such operators to be gauge invariant, such as
\[
 \text{Tr}(\sigma\mp i\rho)^j\,.
\]
The local operators made of $\sigma\mp i\rho$ here, on the other hand,
are not gauge invariant, but they are allowed because the gauge symmetry
is broken at the poles because of the defects.

An important special case is
\begin{equation}
 {\boldsymbol\eta}^\north = {\boldsymbol\eta}^\south =
 \Big(
 \underbrace{\rule[-2.8mm]{0mm}{2mm}0,\ldots,0}_{n_0\text{ times}}\,,\,
 \underbrace{\frac1K,\ldots,\frac1K}_{n_1\text{ times}}\,,\,
 \underbrace{\frac2K,\ldots,\frac2K}_{n_2\text{ times}}\,,\,\cdots \Big),
\end{equation}
which can be understood also as an orbifold $S^2/\mathbb Z_K$ breaking
$U(N)$ to $\mathbb L\equiv\prod U(n_i)$. The one-loop determinant
(\ref{Zvwd}) for this case is nothing but the determinant for the
gauge group $\mathbb L$ in the absence of defects.

\paragraph{Example 2: SQCD.}

As an application of our formalism, let us work out the defect
correlators for the $U(N)$ SQCD with $N_\text{F}$ fundamental and $N_\text{A}$
anti-fundamental matters. As in the case of SQED, we denote the masses
and $\RV$-charges of the $N_\text{F}$ fundamentals as
$m_j=\mu_j+i\rch_j$, and those of the $N_\text{A}$ anti-fundamentals as
$\tilde m_j=\tilde\mu_j+i\tilde\rch_j$.

We begin by summarizing the result for SQED to introduce some
notations. For $|z|<1$ the contour of $a$-integral can be closed in the
lower half plane, and we obtained the formula
\begin{eqnarray}
 \langle V_{\eta^\north}V_{\eta^\south}\rangle^\text{\smaller SQED}
 (z,\bar z) &=&
 \sum_s
 \int\frac{da}{2\pi}\,z^{\eta^\north+ia-s}\,\bar z^{\eta^\south+ia+s}
 \cdot Z^\text{\smaller SQED}_\text{ch}(s,a,\eta^\north,\eta^\south)
 \nonumber \\ &=&
 \sum_{j=1}^{N_\text{F}}
 Z_j\cdot
 F_j(z,\eta^\north)\cdot
 \overline{F}_j(\bar z,\eta^\south),
\label{SQED2}
\end{eqnarray}
where
\begin{eqnarray}
&& Z_j ~\equiv~(-1)^{[\eta^\north]_{\rch_j}}
 \prod_{i\ne j}^{N_\text{F}}(-1)^{[\eta^\north]_{\rch_i}}
 \frac{\Gamma(-im_i+im_j)}{\Gamma(1+im_i-im_j)}
 \prod_{i=1}^{N_\text{A}}(-1)^{[-\eta^\south]_{\tilde\rch_i}}
 \frac{\Gamma(-i\tilde m_i-im_j)}{\Gamma(1+i\tilde m_i+im_j)},
\nonumber \\
&&
F_j(z,\eta) ~\equiv~
z^{-im_j}\,
 F^{[j]}_\text{vortex}
 \left((-1)^{N_\text{F}}z\,;\,\eta,m_i,\tilde m_i \right),
\nonumber \\
&&
\overline{F}_j(\bar z,\eta) ~\equiv~
 \bar z^{-im_j}
 F^{[j]}_\text{vortex}
 \left((-1)^{N_\text{A}}\bar z\,;\,\eta,m_i,\tilde m_i \right)\,,
\end{eqnarray}
and $F^{j}_\text{vortex}$ has been defined in (\ref{Fv}).
In the second line of (\ref{SQED2}), the dependence on the mass
parameters $\mu_i,\tilde\mu_i$ or the $\RV$-charges $q_i,\tilde q_i$ have
been suppressed for later convenience.

To write down the defect correlators in the $U(N)$ SQCD, we first recall
the trick to rewrite the one-loop determinant of the vector multiplet by
differentiations.
\begin{eqnarray}
\lefteqn{
 z^{\text{Tr}(\boldsymbol\eta^\north+i\boldsymbol a-\boldsymbol s)}\,
 \bar z^{\text{Tr}(\boldsymbol\eta^\south+i\boldsymbol a+\boldsymbol s)}\cdot
 \prod_{a<b,\,\eta_{ab}^\north=0\hskip-7mm}(s_{ab}-ia_{ab})
  \prod_{a<b,\,\eta_{ab}^\south=0\hskip-8mm}(s_{ab}+ia_{ab})
} \nonumber \\ &=&
 \prod_{a<b,\,\eta_{ab}^\north=0\hskip-7mm}
 (z_b\partial_{z_b}-z_a\partial_{z_a})
  \prod_{a<b,\,\eta_{ab}^\south=0\hskip-8mm}
 (-\bar z_b\partial_{\bar z_b}+\bar z_a\partial_{\bar z_a})
 \left[
 \prod_{a=1}^N
 z_a^{(\eta^\north+ia-s)_a}
 \bar z_a^{(\eta^\south+ia+s)_a}
\right]_{\substack{z_a\to z\\ \bar z_a\to\bar z}}\,.
\end{eqnarray}
The defect correlator in the SQCD can therefore be obtained by first
acting the above differential operator onto the correlator of $U(1)^N$ theory,
\[
 \prod_{a<b,\,\eta_{ab}^\north=0\hskip-7mm}
 (z_b\partial_{z_b}-z_a\partial_{z_a})\cdot
  \prod_{a<b,\,\eta_{ab}^\south=0\hskip-8mm}
 (-\bar z_b\partial_{\bar z_b}+\bar z_a\partial_{\bar z_a})
 \prod_{a=1}^N\langle V_{\eta^\north_a}V_{\eta^\south_a}
 \rangle^\text{\smaller SQED}(z_a,\bar z_a),
\]
and then setting the $N$ FI-theta parameters to be all equal,
\[
z_a~\to~
(-1)^{N-1}z\equiv\hat z. 
\]
We also need to sum over the Weyl images of
$\boldsymbol\eta^\north, \boldsymbol\eta^\south$ with the weight
$(-1)^{\sigma(\boldsymbol\eta^\north)+\sigma(\boldsymbol\eta^\south)}$.
In the following we write down the formulae for defect correlators
explicitly for two special cases. The generalization is straightforward.

\paragraph{Case 1.} One of the simplest special cases is when
$\boldsymbol\eta^\north, \boldsymbol\eta^\south$ are both nonzero but
proportional to the identity, namely $\eta^\north_a=\eta^\north$ and
$\eta^\south_a=\eta^\south$ for all $a$. Since $\boldsymbol\eta^\north$
and $\boldsymbol\eta^\south$ are Weyl-reflection invariant, we only need
to take the differentiation of the product of SQED correlators.
\begin{eqnarray}
\langle
 V_{\boldsymbol\eta^\north}V_{\boldsymbol\eta^\south}
 \rangle^\text{\smaller SQCD}
 &=&
 \frac{(-1)^{\frac{N(N-1)}2}}{N!}
 \prod_{a<b}^N
 |z_b\partial_{z_b}-z_a\partial_{z_a}|^2
 \sum_{j_a=1}^{N_\text{F}}
 \prod_{a=1}^N Z_{j_a}
 F_{j_a}(z_a,\eta^\north)
 \overline{F}_{j_a}(\bar z_a,\eta^\south)
 \Bigg|_{z_a\to \hat z}
 \nonumber \\
&=& (-1)^{\frac{N(N-1)}2}\sum_{\{j_a\}}\,\prod_{a=1}^NZ_{j_a}
 \cdot \text{det}\left[F_{j_a}^{[b]}(\hat z,\eta^\north)\right]
 \cdot \text{det}\left[F_{j_a}^{[b]}(\bar{\hat z},\eta^\south)\right]\,.
\label{dfc1}
\end{eqnarray}
Here the determinant is that of $N\times N$ matrices with the $(a,b)$-th
entry
\begin{equation}
F_{j_a}^{[b]}(z,\eta)\equiv (z\partial_z)^{b-1}F_{j_a}(z,\eta)\,.
\end{equation}
The summation in (\ref{dfc1}) is thus over the sets of $N$ different
integers chosen from $\{1,\cdots,N_\text{F}\}$.

\paragraph{Case 2.} Another special case is when
$\boldsymbol\eta^\north,\boldsymbol\eta^\south$ are both generic
diagonal matrices with no degeneration of eigenvalues. Then the one-loop
determinant of the vector multiplet is essentially
$(-1)^{\sigma(\boldsymbol\eta^\north)+\sigma(\boldsymbol\eta^\south)}$.
There is no need of differentiation, but we need to sum over all the
Weyl images of $\boldsymbol\eta^\north$ and $\boldsymbol\eta^\south$.
We thus obtain
\begin{eqnarray}
\langle
 V_{\boldsymbol\eta^\north}V_{\boldsymbol\eta^\south}
 \rangle^\text{\smaller SQCD}
 &=&
 \frac1{N!}
 \sum_{\pi,\bar\pi\in S_N}
 \text{sgn}(\pi)\cdot\text{sgn}(\bar\pi)\cdot
 \sum_{j_a=1}^{N_\text{F}}
 \prod_{a=1}^N Z_{j_a}
 F_{j_a}(\hat z_a,\eta^\north_{\pi(a)})
 \overline{F}_{j_a}(\bar{\hat z}_a,\eta^\south_{\bar\pi(a)})
 \nonumber \\
&=& \sum_{\{j_a\}}\,\prod_{a=1}^NZ_{j_a}
 \cdot \text{det}\left[F_{j_a}(\hat z,\eta_b^\north)\right]
 \cdot \text{det}\left[F_{j_a}(\bar{\hat z},\eta_b^\south)\right]\,.
\label{dfc2}
\end{eqnarray}
Again, the summation is over the set of $N$ different integers $\{j_a\}$.
Note also that the matrix elements $F_{j_a}(z,\eta_b)$, as a functions of
$\eta_b$, are locally constant. If $\eta^\north_b$ and $\eta^\north_c$ are close, we would have
\begin{equation}
 [\eta^\north_b]_{\rch_a}=[\eta^\north_c]_{\rch_a}\quad(a=1,\cdots,N_\text{F}),
 \qquad
 [-\eta^\north_b]_{\tilde\rch_a}=[-\eta^\north_c]_{\tilde\rch_a}\quad(a=1,\cdots,N_\text{A}),
\end{equation}
which would imply $F_{j_a}(\hat z,\eta^\north_b)=F_{j_a}(\hat z,\eta^\north_c)$ for all $a$. The first determinant in (\ref{dfc2}) would then vanish because the
$b$-th and the $c$-th columns of the matrix are the same. Especially, for $K=1$ the integer-valued function $[\eta]_\rch$ becomes $\rch$-independent, and the function $F_j(z,\eta)$ depends on $\eta$ only through the integer-valued functions $\lceil\eta\rceil$ and $\lfloor\eta\rfloor$. The correlator (\ref{dfc2}) then vanishes in most cases: for example, if $N_\text{A}=0$ and the $N_\text{F}$ quarks all satisfy the same boundary condition, the determinants in (\ref{dfc2}) all vanish trivially.

\section{Concluding Remarks}\label{sec:concl}

The relation between the gauge theories on vortex defect backgrounds and
orbifold backgrounds is useful in studying the physics of the defects using
path integral formalism. This approach should be applicable in higher
dimensions as well. Similar idea have been employed in the analysis of
surface defects in certain 4D ${\cal N}=2$ gauge theories
\cite{Nawata:2014nca}.

Our results for non-abelian vortex defects imply there may be a dual
description of the defects in terms of local operator insertions.
Some defects in higher dimensions are known to have dual ``electric''
description, namely as low-dimensional field theories on defects
interacting with the fields in the bulk
\cite{Gukov:2014gja,Gomis:2014eya,Assel:2015oxa}.
It is interesting to explore other possible descriptions for the
point-like vortex defects in 2D gauge theories.

\section*{Acknowledgments}

The author thanks Francesco Benini, Heng-Yu Chen, Kentaro Hori and
Tsung-Hsuan Tsai for valuable discussions. He also thanks Yongbin Ruan
and Kentaro Hori for the invitation to the workshop on Non-Abelian
gauged linear sigma model and geometric representation theory in Beijing
where he had useful discussions with the participants.

The author also thanks Sungjay Lee and Takuya Okuda for discussions and collaboration on the related project \cite{Hosomichi:2017dbc} which led to a major revision of this article.

\end{document}